\documentclass[sigplan,screen]{acmart}

\usepackage{multirow} 
\usepackage{tabularx} 
\usepackage{makecell}

\usepackage{subfigure}
\usepackage{graphicx}

\AtBeginDocument{%
  }

\setcopyright{acmlicensed}
\copyrightyear{2025}
\acmYear{2025}
\acmDOI{} 

\acmConference[DIS '25]{ACM Designing Interactive Systems Conference (DIS)}{July 5--9, 2025}{Funchal, Madeira, Portugal}

\begin{document}

\title{Design Probes for AI-Driven AAC: Addressing Complex Communication Needs in Aphasia}



\author{Lei Mao}
\affiliation{%
   \institution{University of Maryland}
  \city{College Park, MD}
  \country{USA}}
\email{maomao@umd.edu}

\author{Jong Ho Lee}
\affiliation{%
  \institution{University of Maryland}
  \city{College Park, MD}
  \country{USA}}
\email{jlee29@umd.edu}
\orcid{0009-0004-0570-7475}

\author{Yasmeen Faroqi Shah}
\affiliation{%
  \institution{University of Maryland}
  \city{College Park, MD}
  \country{USA}}
\email{yfshah@umd.edu}

\author{Stephanie Valencia}
\affiliation{%
  \institution{University of Maryland}
  \city{College Park, MD}
  \country{USA}}
\email{sval@umd.edu}

\renewcommand{\shortauthors}{Mao et al.}

\begin{abstract}


AI offers key advantages such as instant generation, multi-modal support, and personalized adaptability—potential that can address the highly heterogeneous communication barriers faced by people with aphasia (PWAs). We designed AI-enhanced communication tools and used them as design probes to explore how AI's real-time processing and generation capabilities—across text, image, and audio—can align with PWAs’ needs in real-time communication and preparation for future conversations respectively. Through a two-phase “Research through Design” approach, eleven PWAs contributed design insights and evaluated four AI-enhanced prototypes. These prototypes aimed to improve communication grounding and conversational agency through visual verification, grammar construction support, error correction, and reduced language processing load. Despite some challenges, such as occasional mismatches with user intent, findings demonstrate how AI’s specific capabilities can be advantageous in addressing PWAs’ complex needs. Our work contributes design insights for future Augmentative and Alternative Communication (AAC) systems.

\end{abstract}

\ccsdesc[500]{Human-centered computing~Interaction design process and methods}
\ccsdesc[500]{Human-centered computing~Accessibility}
\ccsdesc[100]{Computing methodologies~Artificial intelligence}

\keywords{Aphasia; Augmentative and Alternative Communication; Artificial Intelligence; Research through Design; Language Impairments}



\maketitle

\section{Introduction}

Aphasia, an acquired communication disorder caused by cerebral lesions, affects one-third of stroke survivors~\cite{dickey2010incidence}. People with aphasia (PWAs) face deficits in speech, comprehension, writing, and reading~\cite{asha_site}. Their symptoms and severity are highly heterogeneous, varying from word-finding or pronunciation problems to an inability to recount events~\cite{poirier2024communication}.

People with Aphasia (PWAs) rely on Augmentative and Alternative Communication (AAC) to express themselves~\cite{lasker2008aphasia}. This includes a range of tools, from gestures and low-tech aids (e.g., picture boards, photos) to high-tech devices that allow users to select words or images and generate speech or text~\cite{asha_site,beukelman1998augmentative}.

This study is motivated by understanding how AI’s specific capabilities can be advantageous in addressing PWAs’ complex communication needs. In particular, by allowing participants to interact with four prototypes as design probes, we identified shortcomings in existing features and interaction mechanisms, while also facilitating collaborative ideation to guide the design of future AI-enhanced AAC systems.

Traditional high-tech AAC devices are made up of thousands of vocabulary sets that are difficult to search~\cite{nikolovaBetterVocabulariesAssistive2009, nikolova2010click}. Some studies used contextual information to trigger vocabulary suggestions~\cite{kaneWhatWeTalk2012}, but relied on pre-programmed phrases, limiting usefulness in dynamic contexts. Other approaches used algorithms to predict words without pre-programming~\cite{wandmacher2008methods, wandmacherSibylleAssistiveCommunication2008}, but the prediction was limited to the initial user input, posing a challenge for PWAs who are unable to recall words. Additionally, AAC devices often rely on a single modality, such as text~\cite{williamsCostTurningHeads2016} or images~\cite{allen2007design}, and focus on basic communication needs while neglecting users’ desires for more nuanced self-expression~\cite{nikolova2010click}, and broader social needs~\cite{ibrahim2018design}. Advancements in machine learning and artificial intelligence (AI) create new design opportunities for AAC. Computer vision and Large Language Models have been used to identify objects and suggest phrases for ALS~\cite{kane2017let},  generate symbolic boards from photos for autism \cite{fontana2022dAACAutomatedVocabulary}, and create full sentences from user-provided keywords ~\cite{valencia2023less}. However, since some of these systems are not specifically designed for PWAs, it remains unclear how to adapt AI to PWAs' heterogeneous communication needs.

Our work presented a design probe study of four AI-enhanced AAC to explore how AI's unique capabilities, such as instant generation, multimodal support, and personalized adaptability, can support PWAs. Our study was a two-phase Research through Design (RtD) approach involving 11 participants with Aphasia. The first phase consisted of exploratory interviews with PWAs (N = 7) to understand their expectations for AI. We used insights from these interviews to inform the design of four multimodal and AI-enhanced probes. Each probe addressing specific challenges identified in the interviews. The second phase consisted of interactive evaluations with PWAs (N = 8) to validate the design concepts. Using thematic analysis and inductive coding, we surfaced themes on how novel features from AI might meet the needs of PWAs, and identified areas for improvement and innovation for future AAC systems.

Our contributions include:
\begin{enumerate}
    \item Empirical studies exploring design opportunities for AI to support individuals with aphasia in communication.
    \item Four AI-enhanced system designs were designed to support users during real-time communication and to prepare for future conversations.
    \item Design insights for future Augmentative and Alternative Communication for PWA.
\end{enumerate}

\section{Related Work}

\subsection{High-Tech and AI Solutions for Aphasia}

Aphasia can often cause significant barriers to effectively expressing thoughts and preferences, often impacting an individual's social interactions and daily life ~\cite{spaccavento2013quality}. Multidisciplinary research across computer science, psychology, and speech-language pathology has explored how technology can be designed to improve the quality of life for PWAs~\cite{moffattAphasiaProjectDesigning2004}.

Recent advances in technology, particularly language models, enable faster vocabulary access by suggesting relevant next words for AAC users ~\cite{kim2009context, clark2006context, nikolovaBetterVocabulariesAssistive2009, nikolova2010click}. However, word prediction still relies heavily on initial user input, which is a challenge for PWAs with word retrieval difficulties. To address this, recent studies have explored adaptive systems that suggest words based on contextual cues such as location, past vocabulary, and conversation partner, reducing the burden of word retrieval ~\cite{patel2007enhancing, kim2009context, kaneWhatWeTalk2012, hossain2018using, demmans2012towards}. Recent AI innovations further improve communication efficiency and autonomy, including using computer vision in AAC to identify objects and suggest phrases for ALS~\cite{kane2017let},  generating symbolic boards from photos for autism \cite{fontana2022dAACAutomatedVocabulary}, and integrating LLMs into AAC systems to create sentences from user-provided keywords ~\cite{valencia2023less}. Despite the increased interest in the subject, relatively few works have investigated how AI-based technologies are accessible for PWA, and focused mostly on other AAC users (e.g., AAC users with motor impairments).

Research has also concentrated on improving PWAs’ rehabilitation and social participation ~\cite{spaccavento2013quality}. Researchers have advanced language therapy by supporting long-term recovery with rehabilitation games~\cite{hymesDesigningGameBasedRehabilitation2021}, improving storytelling skills by integrating personal photographs with calendars \cite{woudstraSnapshotDiarySupport2011}, and improving communication skills through mixed real-virtual environments ~\cite{stapletonTransformingLivesStory2014, galliersExperiencingEVAPark2017}. 
Additionally, machine learning has been applied to predict recovery from aphasia, and multimodal digital pens have been developed to assist speech therapists. ~\cite{laiExplorationMachineLearning2021, guMachineLearningApproach2020, piper2011write}. Technologies to address PWAs' reduced social participation ~\cite{spaccavento2013quality} include applications for meal ordering ~\cite{obiorahDesigningAACsPeople2021},  tools for managing digital photographs ~\cite{allenFieldEvaluationMobile2008}, multimedia representations of unknown words~\cite{maOnlineMultimediaLanguage2010}, discreet and wearable AAC devices that offer support with less stigma~\cite{curtis2023watch, williamsDesigningConversationCues2015, williamsCostTurningHeads2016}, and AphasiaWeb, a social network illustrating to foster community and reduce isolation among PWAs~\cite{millerAphasiaWebSocialNetwork2013}. 

Despite recent developments, it remains unclear what AI qualities can effectively address PWAs’ complex communication barriers in the day-to-day, whether at the linguistic or social level. 

\subsection{Research through Design with People with Aphasia}

Research through Design (RtD) is a Human-Computer Interaction (HCI) approach that addresses research questions while producing tangible design outcomes through user-centered design practices~\cite{zimmermanResearchDesignMethod2007, godinAspectsResearchDesign2014}. It is especially valuable when traditional usability studies are insufficient for exploring future concepts~\cite{kankainenUCPCDUsercenteredProduct2003, gaverDesignCulturalProbes1999}. 

The RtD approach typically results in prototypes featuring innovative interaction paradigms by involving end-users in the design process~\cite{zimmermanResearchDesignMethod2007}. Integrating PWAs in this approach presents unique challenges, including communication barriers, and cognitive fatigue~\cite{wilsonCodesignPeopleAphasia2015, galliersWordsAreNot2012}. However, prior research has successfully involved PWAs throughout the design process, empowering them and enhancing their autonomy~\cite{wilsonCodesignPeopleAphasia2015}. For example, Hymes and colleagues engaged PWAs in designing three hybrid digital-analog games~\cite{hymesDesigningGameBasedRehabilitation2021}, Obiorah and colleagues co-designed three AAC apps specifically for restaurant dining, addressing a critical social need~\cite{obiorahDesigningAACsPeople2021}, Kane and colleagues collaborated with five PWAs to develop and evaluate context-aware AAC systems~\cite{kaneWhatWeTalk2012, kane2017let}, and Curtis and colleagues involved PWAs in creating discreet wearable AAC solutions~\cite{curtis2022wearable, curtis2023watch}. Other efforts have focused on fostering creative expression among PWAs through structured co-design processes, such as comic and poetry creation, and digital content development~\cite{neate2019empowering, tamburro2020accessible,neate2020creatable}. These initiatives highlight the potential of RtD in developing targeted AI-driven technological solutions with and for PWAs.

\subsection{Design research with Generative AI tools}

In user-centered design, generating innovative ideas is defined as “building the right thing,” while prototyping and iteration are ``building it right''~\cite{buxton2010sketching}. Many technological products fail not because of technical infeasibility but due to a lack of consideration for real user needs during the design process~\cite{buxton2010sketching, cooper1999inmates}, and similar failures have occurred in AI projects~\cite{weiner2022ai, yildirim2023investigating}. However, human-centered AI design presents two key challenges. The first involves the unique properties of AI as a design material, such as explainability~\cite{ehsan2021operationalizing}, interpretability~\cite{abdul2018trends}, trust~\cite{yin2019understanding}, user control~\cite{louie2020novice}, and managing user expectations~\cite{kocielnik2019will}. The second challenge is designers’ understanding of AI capabilities. Underestimating AI limits conceptualization and exploration of design opportunities, while overestimating AI leads to unfeasible and impractical design ideas~\cite{dove2017ux, jameson2007adaptive}.

Recent studies have begun exploring methods for better integrating AI into design. For example, Vaithilingam and colleagues proposed to focus on AI's three key functions in design: dynamic dialogue, constructive negotiation, and sustained motivation~\cite{vaithilingam2024imagining}; Yildirim translated AI jargon into action verbs like “create” and “recommend,” making AI capabilities more accessible to designers~\cite{yildirim2022experienced, yildirim2023investigating}; Major technology companies such as Microsoft ~\cite{amershi2019guidelines}, Google~\cite{google_pair_2021}, and Apple~\cite{apple_hig_2023} have developed specific AI and Machine Learning design guidelines. 

HCI designers can gain deeper insights into AI’s capabilities and limitations through real-world project development~\cite{buxton2010sketching, redstrom2005technology,wiberg2014methodology}. As a result, research increasingly explores how AI can address accessibility needs. For instance, NLP has been integrated into AAC systems to reduce users’ typing time and~\cite{valencia2023less, valencia2024compa}, and into intelligent writing tools to assist individuals with dyslexia in composing emails~\cite{yang2019sketching, goodman2022lampost}. Image generation has been used to create pictures, while LLMs explain pictures to blind and low-vision creators~\cite{huh2023genassist}. Building on this, we used fully functional prototypes to explore how generative AI can address PWAs’ complex communication needs, informing future AAC system designs.

\section{Phase One: Exploratory Interviews}


We started our research-through-design process by opening a formative inquiry with seven individuals with aphasia. This first phase was aimed at better understanding current opportunities and challenges for AI-enhanced communication support.

\begin{table*}
  \caption{Participant Demographics. Participants self-identified their perceived abilities on speaking, listening, reading, and writing comprehension, where they chose from the following options: Unable, a few words, some phrases, complete sentences, conversations, no difficulty.}
  \label{tab:participants}
  \footnotesize
  \begin{tabular}{ c m{5em} l m{2em} m{7em} m{7em} m{7em} m{7em} }
    \toprule
    \textbf{PID} & \textbf{Phase \newline Attended} & \textbf{Gender} & \textbf{Age} & \textbf{Speaking} & \textbf{Listening} & \textbf{Reading} & \textbf{Writing} \\
    \midrule
    AP1 & Phase 1, \newline Phase 2 & Male & 59 & Have \newline conversations & Understand \newline conversations & Some phrases & A few words\\
    \midrule
    AP2 & Phase 1 & Female & 75 & Have \newline conversations & Understand \newline conversations & Complete \newline sentences & A few words\\
    \midrule
    AP3 & Phase 1, \newline Phase 2 & Female & 72 & Have \newline conversations & Understand \newline conversations & No difficulty & Some phrases\\
    \midrule
    AP4 & Phase 1, \newline Phase 2 & Female & 62 & A few words & Complete \newline sentences & Complete \newline sentences & Complete \newline sentences\\
    \midrule
    AP5 & Phase 1 & Male & 46 & Some phrases & No difficulty & A few words & A few words\\
    \midrule
    AP6 & Phase 1 & Female & 76 & Have \newline conversations & No difficulty & No difficulty & Paragraphs\\
    \midrule
    AP7 & Phase 1, \newline Phase 2 & Female & 51 & Complete \newline sentences & Complete \newline sentences & Some phrases & A few words\\
    \midrule
    AP8 & Phase 2 & Male & 50 & Complete \newline sentences & Complete \newline sentences & Complete \newline sentences & Complete \newline sentences\\
    \midrule
    AP9 & Phase 2 & Female & 47 & Have \newline conversations & No difficulty & Paragraphs & Paragraphs\\
    \midrule
    AP10 & Phase 2 & Male & 51 & Some phrases & Some phrases & Some phrases & Some phrases\\
    \midrule
    AP11 & Phase 2 & Male & 57 & Have \newline conversations & No difficulty & No difficulty & Complete \newline sentences\\
    \bottomrule
  \end{tabular}
\end{table*}

\subsection{Method}

Seven PWAs (Table~\ref{tab:participants}) were recruited through an aphasia support groups across the United States. Participants (mean age: 63; range: 46–76) included both genders and varied in communication challenges and technology use. Each of our participants received \$25 for a one-hour Zoom interview. AP5 participated with his mother as a conversation partner, while the other six participants attended independently.

Participants received interview questions three days in advance. The semi-structured interviews began with an explanation of the study’s objectives, followed by questions about their aphasia recovery, communication challenges, and strategies for overcoming them. Questions such as ``What gets in your way when speaking?'' and ``What can help you in those situations?'' guided the discussion. Storyboards were then introduced to explain AI concepts and inspire design ideas.  We also used storyboards to discuss the possible usefulness of specific AI capabilities across diverse scenarios. The topics included: (1) Image-to-text translation to facilitate understanding medical prescriptions, (2) Speech-to-text to facilitate email writing, (3) Keyword-based sentence generation for ordering coffee, and (4) an AI virtual therapist with whom to practice  to improve language rehabilitation.



Interviews were transcribed and analyzed using thematic analysis~\cite{braun2006using}. Codes were generated from the transcripts by two researchers, which were then organized into overarching themes.

\subsection{Interview Findings}\label{sec:phase1findings}

In this subsection, we report on design insights gathered through our phase one interviews. We first discuss interaction modalities that pertain to PWAs' accessibility needs. We then dive into identified challenges including linguistic barriers, social isolation, and cognitive burdens. 

\subsubsection{Interaction Modality Needs and Expectations.} 


Our participants emphasized the need for accessible, intuitive, and multimodal tools to support daily communication while reducing physical and cognitive effort.

Motor disabilities that might make using both hands to type challenging, commonly impact stroke survivors with Aphasia.  AP1 reported numbness in his right hand after a stroke, restricting his ability to use technology. AP3 mentioned during the interview that she uses a wheelchair, which limits her daily activities. AP4 highlighted this frustration, stating: \textit{“I can no longer drive. My right hand. I can’t do it anymore.”} To reduce physical effort, PWAs often respond to messages with just one or two words or emojis. AP5’s mother explained: \textit{``He doesn’t type. He sends messages using emojis.”} Due to this, we identified that using \emph{\textbf{speech input}} in future interactive probes could additionally increase accessibility for our user group. 



\emph{\textbf{Images and Visual Aids}} are another important feature and useful tool to PWA. When language processing areas in the brain are affected, visual aids become essential for PWAs, helping with both self-expression and understanding. AP7 shared an instance at a grocery store where she relied on images to request a sandwich she couldn’t name. Similarly, AP5 uses a wordless travel book with images for their daily communication. AP1 also relied on images for support: \textit{“I use Google Image… tell my wife… like hot dog.”} Visual aids were also used to assist others in understanding PWAs better. AP3 noted: \textit{“Sometimes it’s hard to explain. Images make things clearer.”}

Due to reading difficulties, PWAs prefer repeated \emph{\textbf{audio playback}} for better comprehension. They avoid fast-paced speech and favor brief, clear, and slow communication. AP1 explained: \textit{“I need slow… Not like a cartoon, but slow.”} Fast-paced TV speech can make focusing a challenge. AP5 also mentioned experiencing frustration when being rushed to respond: \textit{“It’s frustrating and annoying.”} AP6 suggested text-to-audio conversion with slow, concise speech to reduce information overload: \textit{“After stroke, I need people to speak slowly. It’s hard to understand when they say too much at once.”}

Beyond multimodal interaction, PWAs prefer smartphones over computers for daily communication due to their convenience. AP1 stressed: \textit{“The phone is God.”} Our participants favored simple, intuitive interfaces to avoid complex interactions and information overload. Stability and consistency are also important. AP1 expressed a preference for \textit{“long-term use of one technology”} to avoid frequent changes, saying: \textit{“I don’t like always updating… update very soon… I have to learn.”} Nonetheless, some participants showed strong interest in newer smart language processing features like Grammarly’s \textit{“auto-correct”} (AP1, AP2, AP7) and word prediction in Microsoft Word and email applications (AP1, AP3).

Although some participants did not fully understand AI’s capabilities—AP2 remained confused about \textit{“what AI can do,”} and AP5 and AP6 admitted, \textit{“I never heard about it”}—most expressed a strong desire for AI to help them regain their language abilities. AP2 remarked: \textit{“If my language could come back, I’d give anything.”} Similarly, AP5 shared: \textit{“I hope to speak… like before.”}

\subsubsection{Current Linguistic and Collaborative Efforts}

Retrieving words is especially hard for PWAs. Our participants shared that communicating can take a lot of effort, can cause fatigue and still feel dissatisfying. AP1 described the process of trying to find the right words as running a \textit{“marathon,”} due to the duration it takes and the effort. Although some participant can use simple words, they feel frustrated by their inability to express deeper emotions. AP1 shared, \textit{“I didn’t want to just say ‘I am good’. I wanted to say more.”} PWAs described some collaborative strategies that helped him with word retrieval:

\textbf{Word Completion:} PWAs may recall only part of a word, which is especially true for longer, multi-syllabic words. In these situations, proactive completion by a conversation partner can be effective. AP3 shared, \textit{“When I wanted to say ‘my nephew’s kindergarten graduation ceremony,’ I couldn’t remember how to say ‘graduation.’ I kept repeating ‘grad,’ and then someone helped by saying, ‘Oh, you mean graduation.’ That was helpful.”}

\textbf{Follow-up Questions for Confirmation}: When PWAs struggle to articulate their thoughts, familiar conversation partners often ask questions to clarify their intended meaning which can be helpful. For example, AP2 shared that when she has trouble finding the right word, her husband helps by asking questions that start by ``do you mean..?", like  \textit{``Do you mean chicken?”} Similarly, AP5’s mother explained that due to the severity of AP5's aphasia, she often relies on repeated questioning to understand his needs.

\textbf{Related Words to Trigger Associations}: When PWAs cannot recall a target word, they often use related or associative words to provide contextual clues. For example, AP5 struggled to recall the word \textit{“sugar”} but successfully conveyed his meaning by saying \textit{“coffee,”} which helped his mother understand. Similarly, AP1 explained his strategy of using related words to trigger associations in the listener’s mind when the exact term eludes him. He shared an example involving the word \textit{“apple”}: \textit{“One time it’s like it, it is, or it isn’t. But 2 or 3 times the apple, and then a real apple, not real apple, but a fake apple. And then oh, now, now I got it so.”} In this context, \textit{“fake apple”} refers to related terms that evoke the correct word through association, allowing the listener to infer his intended meaning even when the precise word cannot be recalled.

\subsubsection{Timing: Real-time and Practicing for Conversations}
While these collaborative strategies can facilitate communication, PWAs also face timing issues in \textbf{real-time conversations} that can cause additional complexity.Delays in word retrieval often cause conversational breakdowns and awkward timing. AP6 noted, \textit{“It’s very embarrassing to keep people waiting for a long time while I try to answer.”} Our participants expressed wanting instant support in real-time conversations because \textit{"there are always unpredictable topics in life. You can’t always be prepared for each one" }(AP7). Participants also wanted support when \textbf{prepararing for future conversations}. Because \textit{"unprepared conversations make me nervous"} (AP6). Therefore, AP3 suggested using technology to practice in advance for real-world scenarios: \textit{“I wish my speech therapy was half grammar and half practice for real conversation. Because I still don’t know how to talk to people in a pharmacy or at a cashier.”}

\subsubsection{Invisible Social Isolation and Negative Emotional Consequences}

In addition to linguistic and timing challenges, PWAs face hidden issues like negative emotions and social isolation, especially during real-time interactions. The main difficulty is the disconnect between thoughts and expression, which can cause distress in daily life, particularly when sharing \textit{“important information”} (AP2). AP2 described her fear of making mistakes and the guilt it brings: \textit{“I wanted to meet on Wednesday, but I ended up saying Monday. I’m really afraid of such mistakes. I feel guilty.”}

Our participants shared often experiencing social exclusion due to their speech difficulties. AP3 shared how in public, they are sometimes misunderstood as \textit{“drunk”} or \textit{“crazy”} and face impatience from strangers. AP2 sarcastically noted, \textit{“Store clerks don’t have time to play games with me.”} Referring to the effort sometimes required from a speaking partner to try an understand them. AP7 shared how even in private settings, she is frequently ignored in group conversations. AP7: \textit{“People turn to my husband for answers. I can speak. They just need to wait for me to finish.”} AP2 also mentioned, \textit{“Sometimes they just turn around and leave.”}

These experiences lower self-esteem (AP3, AP7), reduce social interactions (AP2), and worsen language abilities (AP3). Our participants with aphasia often suppress their needs by simplifying their messages. AP5 said that after repeated failed attempts to communicate, he would give up or agree with others, even if it wasn’t what he meant.In general, PWAs expressed wanting to communicate clearly and accurately the first time to avoid misunderstandings and corrections. They also hoped to overcome language barriers in real-time conversations so that they can fully participate without being ignored or excluded.

Additionally, PWAs experience emotional impacts due to inability to describe meaningful past experiences, which often results in a sense of partial loss of self and anxiety about \textit{``losing control over life”} (AP2). They also long for deeper talk beyond the superficial exchanges their language limitations impose. AP1 shared, \textit{“When I call my daughter, I don’t want to just say ‘hi, how is it going?’ I want to talk about other things.”}

\subsubsection{Cognitive Burden, Sense of Control, and Self-Disclosure}

PWAs reported facing cognitive burdens during language processing, particularly with large amounts of input and descriptive output. For comprehension, PWAs shared struggling with long texts and conversations. AP5 noted that long articles and lengthy conversations made him feel \textit{“tired and sleepy.”} A common suggested potential use of AI was to \textit{“break down long texts”} and summarize them into \textit{“key points”} (AP2, AP7). Participants also expressed frustration and cognitive effort when trying to describe events. AP1 described this process as \textit{“running a marathon,”} while AP2 likened it to \textit{“a big chunk of cheese in my brain, but I can only express a small piece of it.”} AP3 repeatedly said, \textit{“I know, but I can’t do it.”} To ease this burden, they often relied on \textit{“photos”} as visual aids (AP3, AP5).


Due to the importance of meaningful conversations, many PWAs prepare for them in advance. AP2, AP4, and AP7 expressed enthusiasm for the \textit{“virtual therapist for conversation rehearsal”} feature mentioned in the storyboard, with AP2 saying, \textit{“I really like this.”} AP6 hoped AI could provide prompts to practice conversations: \textit{“I want an AI. It can ask me questions, adjust the difficulty, and help me restore my vocabulary.”} She emphasized that AI should serve as an \textit{“assistive tool, not a directive tool,”} as she wanted to maintain her ability to express herself. However, AP6 also raised concerns about whether AI could capture the subtle nuances of her communication, asking, \textit{“Can AI capture my subtle nuances, or will it only have very general conversations?”}

\subsubsection{Identified Design Goals}
Based on our phase one findings, we designed four systems to explore how AI can support PWAs’ communication needs, addressing both linguistic and social challenges:

\begin{itemize}
    \item \textbf{Leveraging images to support word retrieval} From a linguistic perspective, System 1 is inspired by the need for follow-up confirmation to support collaborative word retrieval and using images to facilitate communication. Socially, it also aims to reduce guilt, fear of mistakes, and misunderstandings. 
    \item \textbf{Using word associations to construct sentences} System 2 is inspired by the need for contextual association words and grammar construction from a linguistic perspective. Socially, it supports fluent and timely responses in real-time conversations. 
    \item \textbf{Practicing and Refining} System 3 is inspired by the need to practice and prepare for conversation. We can use AI's word association and word suggestion capabilities to enhance collaborative communication from a linguistic perspective, and to encourage richer expression from a social perspective.
    \item \textbf{Personal Storytelling Support} System 4 is inspired by the need to reduce cognitive burden when retelling past experience while also enhancing a users' sense of control over their life, and provide opportunities for conversation practice.
\end{itemize}

We explain these systems in depth in the next section.

\section{Prototype Design}

A crucial component of our study design is the use of four prototypes as design probes. Two of our design probes focus on supporting real-time communication (systems 1 and 2) and two other probes focus on dialogue preparation (systems 3 and 4). All systems allowed text input or voice input, to provide accessible input options.  


\subsection{System 1: Double Check Important Words}

\begin{figure}[h]
  \centering
  \includegraphics[width=0.9\linewidth]{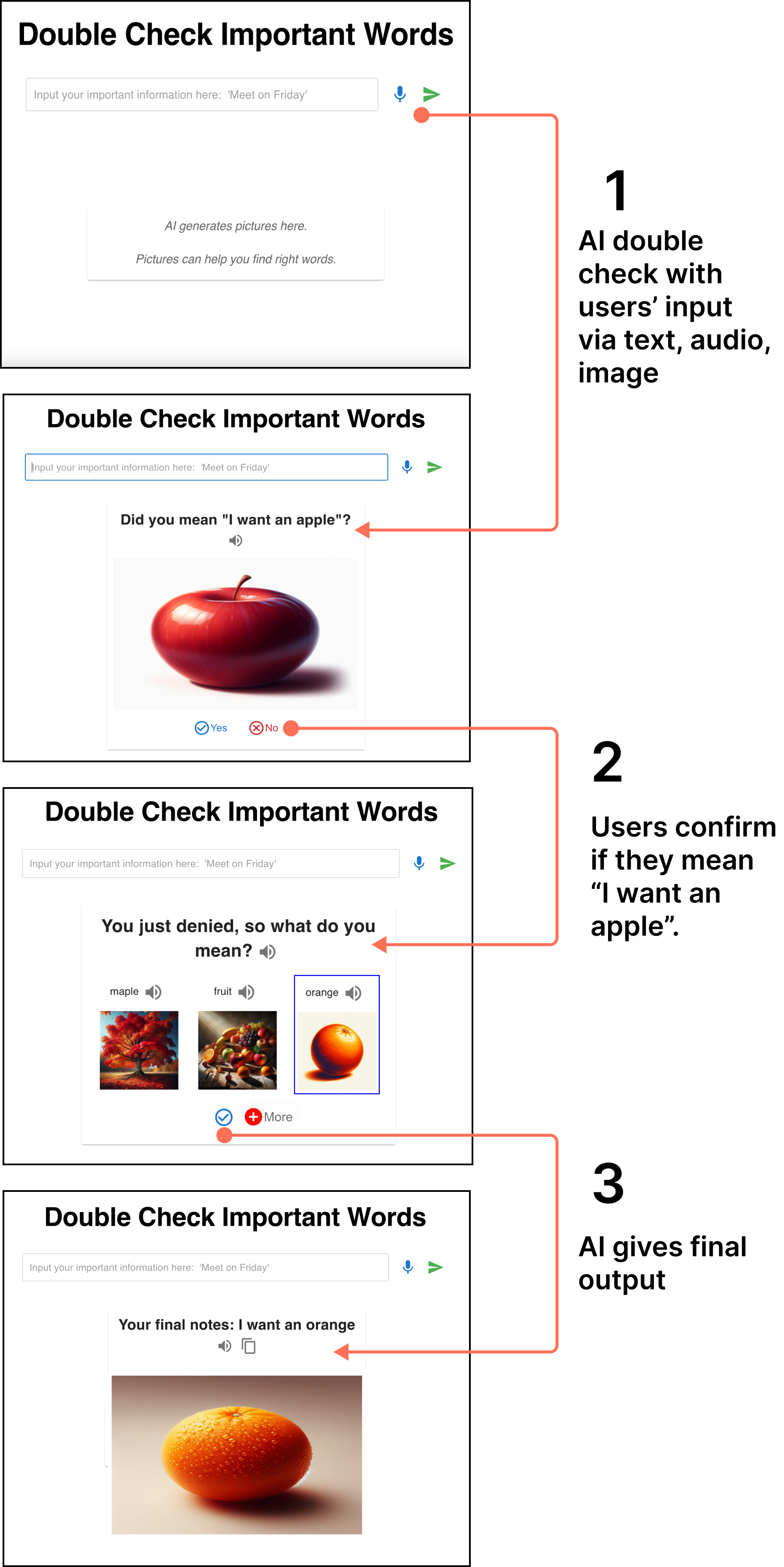}
  \caption{System 1 allows users to check and confirm intended words using AI-generated images. \newline
  \textbf{Step 1:} Users input notes via text or voice. \newline
  \textbf{Step 2:} AI generates an image and asks for confirmation. \newline
  \textbf{Step 3:} Users confirm or correct the AI's output, and the final notes are generated.}
  \label{fig:system1}
  \Description{
    A screenshot of System 1 showing the process where users input notes, AI generates images, and users confirm or correct the output.
  }
\end{figure}


System 1 is designed for users who want to express important needs while ensuring their message is accurate (Figure~\ref{fig:system1}). Users begin by inputting notes—via voice or keyboard—containing the information they wish to convey. The system then uses OpenAI’s DALL.E 3 to generate an image based on the user’s notes (input). This image serves as a visual representation of their intended message. The user is then asked to confirm if the generated image accurately represents their input. If the user selects “yes,” indicating the image aligns with their intended message, the system outputs the original notes as the final message. If the user selects “no,” the process enters a correction phase. In this phase, the system generates three alternative words or phrases based on the original notes and creates three new images reflecting these options. The user reviews these images and selects the one that best matches their intended message. Once confirmed, the system updates and displays the final corrected notes.


\subsection{System 2: Generate Sentences from Keywords}

\begin{figure}[h]
  \centering
  \includegraphics[width=0.9\linewidth]{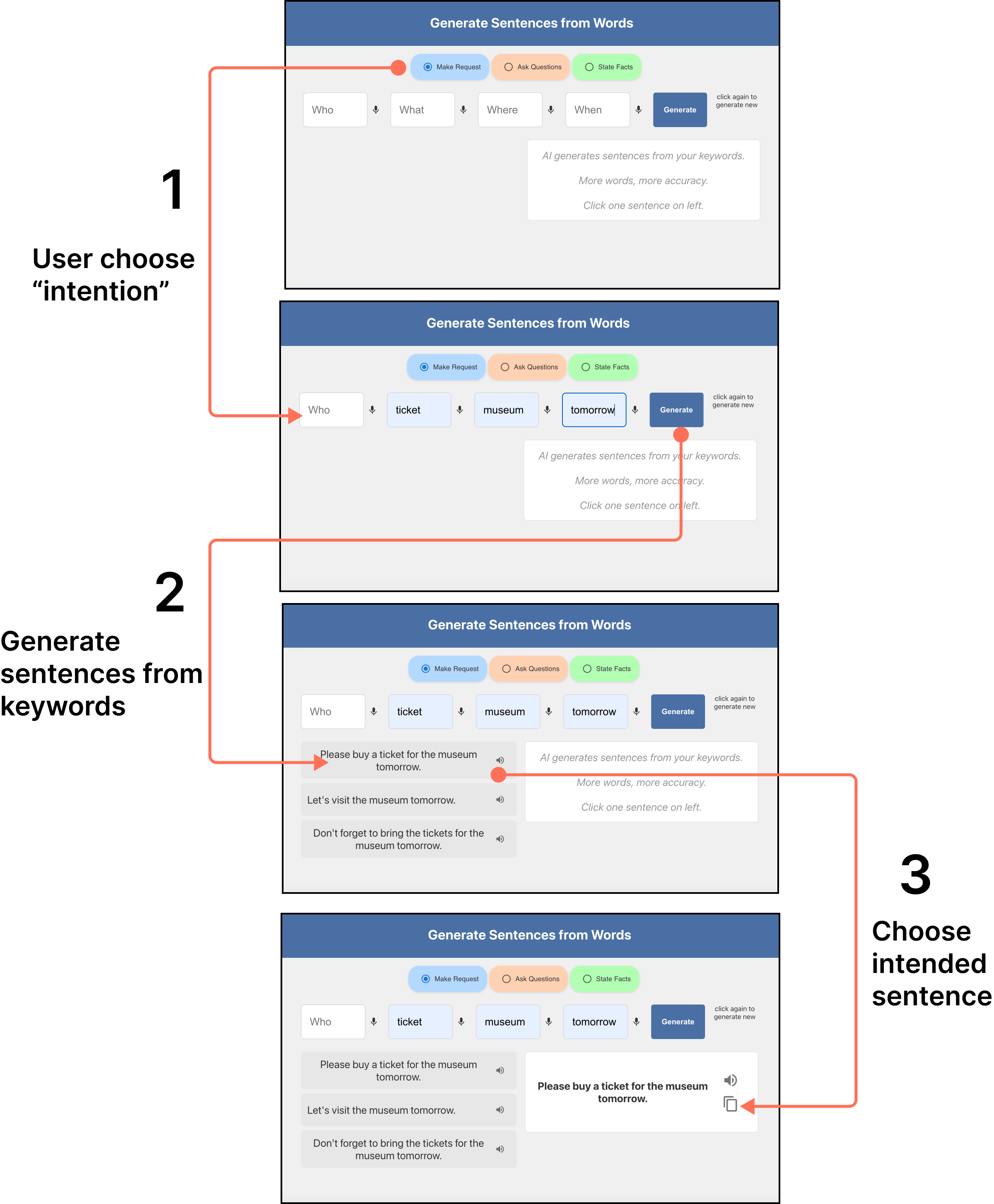}
  \caption{System 2 allows users to generate sentences from keywords and intentions. \newline
  \textbf{Step 1:} Select intention type (e.g., request, question). \newline
  \textbf{Step 2:} Input 2–4 keywords. \newline
  \textbf{Step 3:} Review and confirm AI-generated sentences.}
  \label{fig:system2}
  \Description{
    A screenshot of System 2, showing the process of selecting an intention type, entering keywords, and reviewing AI-generated sentences.
  }
\end{figure}


System 2 is designed to assist users who can recall keywords but struggle with grammar and sentence construction (Figure~\ref{fig:system2}). Users begin by selecting the "intention" button: raising requests, asking questions, or stating facts. Based on the chosen category, the system prompts the user to input 2-4 keywords corresponding to “who,” “what,” “when,” and “where.”Using these keywords, ChatGPT Turbo 3.5 generates three simple sentence options. The user reviews these options, aided by a text-to-speech function, and selects the sentence that best matches their intention. Once the user confirms their choice, the system outputs the selected sentence, ready for communication. This process helps users construct clear and grammatically correct sentences, enabling them to express themselves effectively and engage with others more confidently.


\subsection{System 3: Check Whole Sentences}

\begin{figure}[h]
  \centering
  \includegraphics[width=0.9\linewidth]{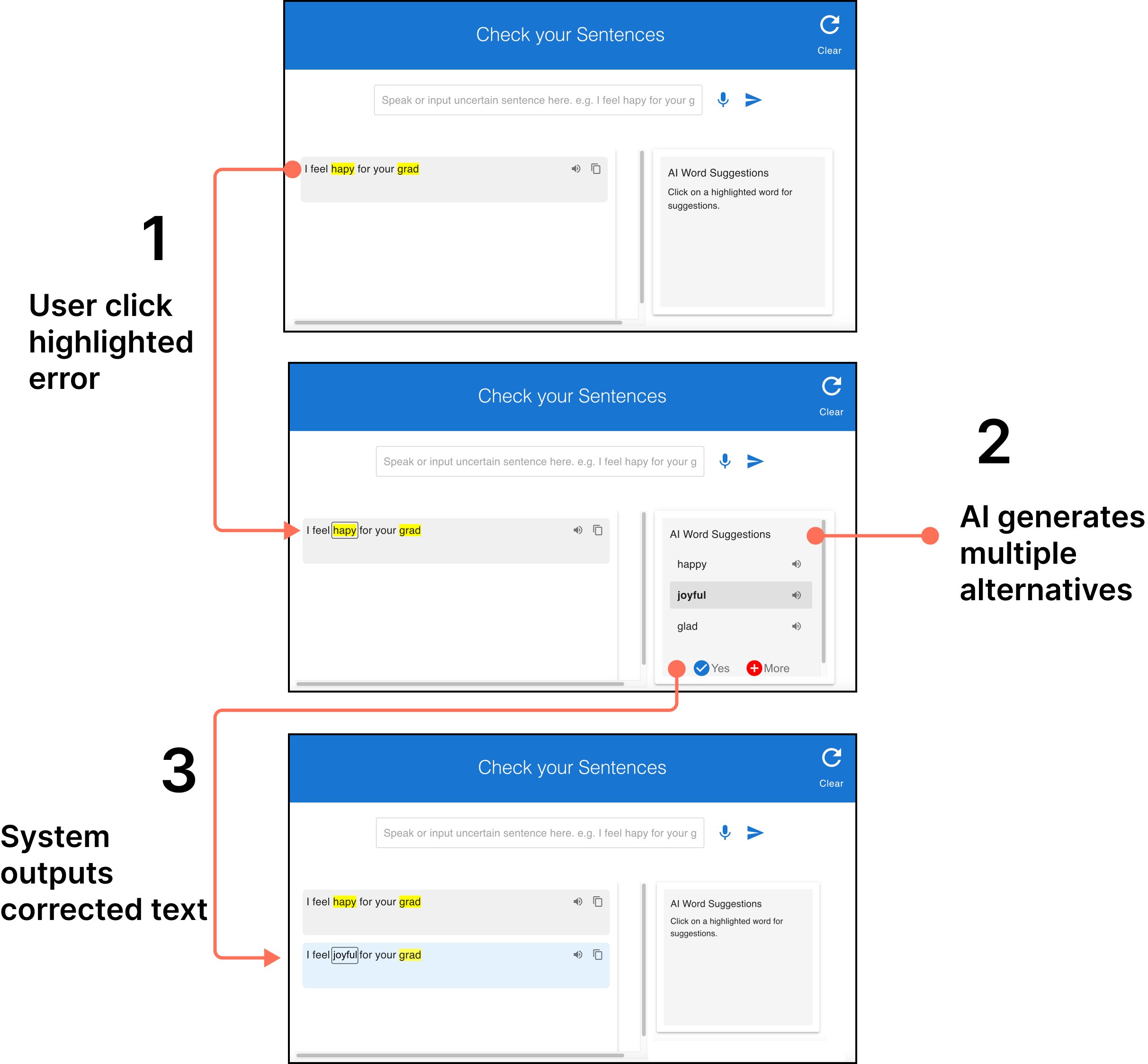}
  \caption{System 3 allows users to check and correct the specificity of whole sentences. \newline
  \textbf{Step 1:} Users input sentences via text or voice. \newline
  \textbf{Step 2:} AI highlights errors (misspellings, incomplete, context errors). \newline
  \textbf{Step 3:} Users select corrections, and the final sentence is generated.}
  \label{fig:system3}
  \Description{
    A screenshot of System 3, showing the process of inputting sentences, AI highlighting errors, and users correcting them.
  }
\end{figure}


System 3 is designed to assist users in recalling words, correcting mispelled words, and enhancing their expression by suggesting synonyms (Figure~\ref{fig:system3}). Users start by inputting a sentence via voice or keyboard, including any uncertain or incomplete words. ChatGPT 3.5 Turbo analyzes the input and identifies potential errors. The system highlights these errors with color codes: yellow for misspelled words, orange for incomplete words, and pink for contextually incorrect words. When a user clicks on a highlighted word, the system offers three replacement suggestions to address the issue. The user selects the most appropriate option, and the system integrates the correction into a revised sentence. This corrected version retains the user’s original intent while addressing errors. The final output is a clear and accurate sentence, ensuring the user’s message is effectively conveyed and reducing misunderstandings.


\subsection{System 4: Share Your Meaningful Experience}

\begin{figure}[h]
  \centering
  \includegraphics[width=0.9\linewidth]{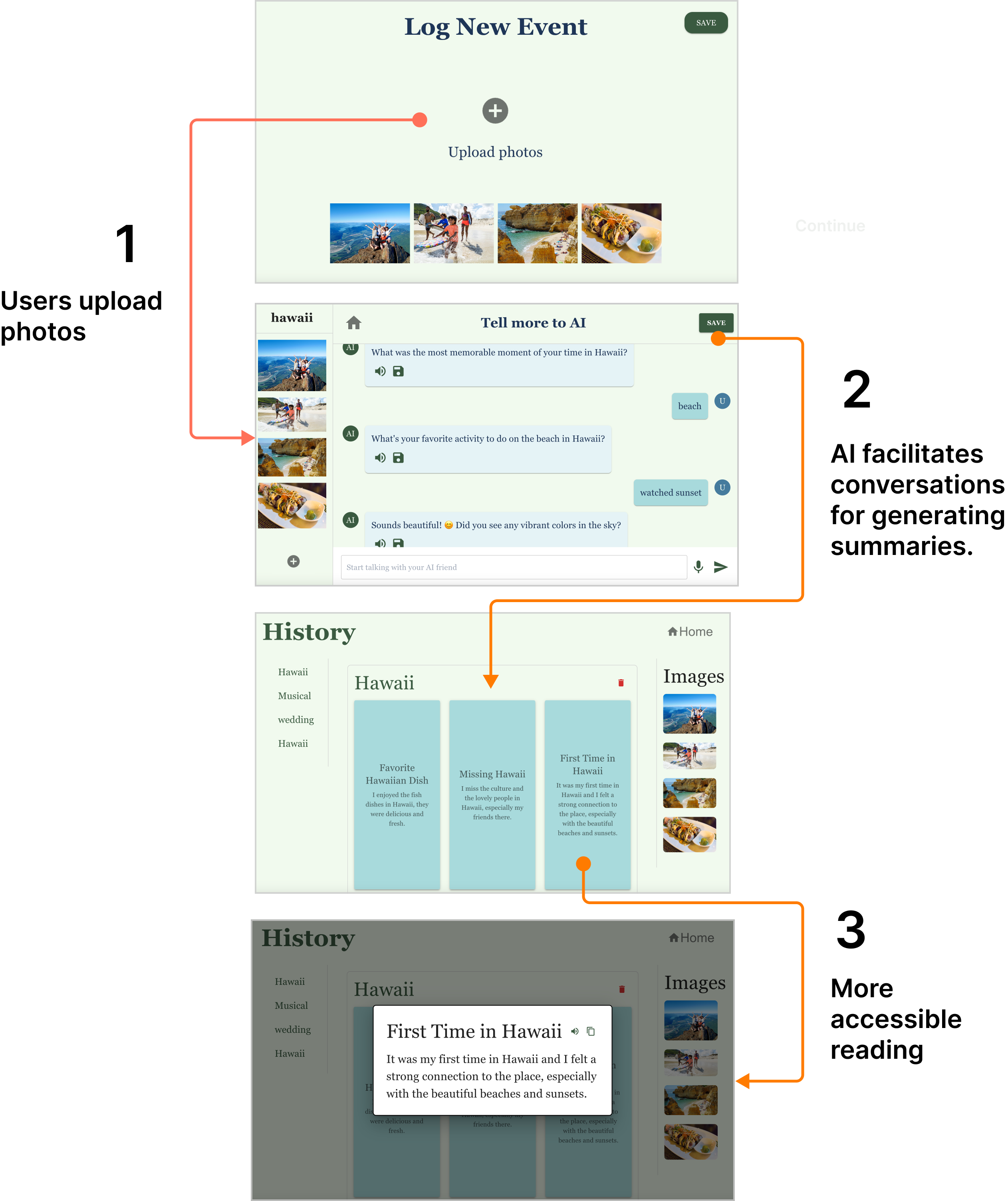}
  \caption{System 4 allows users to create diary entries by engaging in conversations with an AI chatbot. \newline
  \textbf{Step 1:} Users input a diary topic and upload photos. \newline
  \textbf{Step 2:} AI engages in a conversation to collect details. \newline
  \textbf{Step 3:} System organizes the diary entry into subtopics for future review.}
  \label{fig:system4}
  \Description{
    A screenshot of System 4 showing the process of creating a diary entry with user inputs, AI-generated conversations, and organized subtopics.
  }
\end{figure}


System 4 is designed to help users record specific events for future recall, reducing the cognitive burden of organizing speech (Figure~\ref{fig:system4}). Users start by inputting a brief diary topic via speech or keyboard and upload related photos. These photos act as visual aids, enhancing memory and providing context for the recorded event. To ensure privacy during testing, the system displays four pre-stored photos, to showcase functionality to our participants instead of requiring the need of personal photo uploads.

Next, ChatGPT 3.5 Turbo engages the user in a simple dialogue, gathering detailed information about the event. The chatbot asks the participants to tell them about their life events. Based on this conversation, the system categorizes the content and generates relevant subtopics, making the information easier to navigate and recall later. Users can browse their history to view past topics, and selecting a topic reveals its associated subtopics. Clicking on a subtopic displays a conversation summary organized by the system, ensuring users can efficiently retrieve and review event details when needed.


\section{Phase Two: Interactive Evaluation}

For the second phase of this work, we completed an interactive evaluation of our four design probes with PWA over a remote video call. Our goal was to understand what specific AI features were most useful to address PWAs' complex communication needs.

\subsection{Method}

In total, eight PWAs (4 females; mean age = 56, range: 47-72) participated in phase two, with four participants continuing from phase 1. Participants presented varied aphasia-related challenges (Table 1). All participants attended two 60-minute Zoom sessions (120 minutes overall) on consecutive days to evaluate four prototypes: two focused on real-time communication (Day 1) and two on future dialogue preparation (Day 2).

During the interactive evaluation, we first asked users about their current strategies for addressing specific communication barriers, such as, "What do you typically do to speak a full sentence?" Next, we introduced each system, explaining its design goals and usage scenarios using storyboards, followed by researcher-led demonstrations on how to use the system. To help PWAs better understand the different roles of the system, our storyboard follows a cohesive storyline, “Cruise Travel.” The topics of storyboards for Systems 1-4 within this context are: (a) ordering food at a restaurant, (b) buying tickets at a theater, (c) writing a letter to a nephew about the trip, and (d) preparing a travel story to share at a gathering of friends. After each storyboard, participants were asked to: (1) imagine themselves in the storyboard scenarios, and (2) to recall similar real-life challenges to solve using the system. 

Given participants’ age, language impairments, and motor disabilities, researchers navigated the interfaces while participants provided input via speech or Zoom’s chat box. The interface view and all storyboards were shared using screen sharing. This approach ensured participants could focus on evaluating system functionality rather than learning to navigate interfaces. After testing, the questionnaire with semi-structured interviews was conducted, assessing five metrics: goal completion, trust in AI, grounding with AI, the usefulness of specific AI features, and overall system usefulness.

Data collection included session recordings, user inputs, system outputs, and questionnaire responses. Two researchers independently coded transcribed interviews and comments using thematic analysis~\cite{braun2006using}, followed by iterative categorization.

\subsection{Results - System 1: Double-check Important Words}

System 1 uses AI-generated images to help PWAs verify their expressions by providing visual confirmation of their intended meaning.

\subsubsection{Current Strategies for Word Verification}

Our participants shared they confirm the accuracy of their words through multiple methods: visual aids (phone photos, stickers) (AP3), dictionaries (AP4), and pre-written lists (AP7). AP7 shared:  \textit{"I get nervous when I’m out… If I can’t pronounce the word, I show my list to explain what I’m trying to say."} Some participants utilize semantic relationships - opposites (AP5) or synonyms (AP4) - while others engage in iterative verbal attempts to identify words through context, as demonstrated by AP1's search for the word "glass":

\textit{“For me the words are really really…not good at all. So I a mouth a right…figure it out…some kind of thing…and…so called. It’s just…not this one. And then that’s it—a drink, but not a drink, and that’s it…a drink. A glass. Okay? So the glass for me. Oh, the glass. Now I got it. Okay, or whatever it is. So…it’s just a little…weird for me…just…talking and talking only so talking. And here, whatever it is, then it’s oh, now I got it so.”} 

(Researcher's interpretation: For me, the words I use are really not good at all. I usually figure it out as I go. For example, if I’m thinking of something like “a drink,” but not exactly that, I might struggle to find the right word. Then it clicks—“Oh, a glass! That’s it.” It feels a bit weird just talking and figuring things out as I speak - “Oh I got it”.)

\subsubsection{Enhanced Grounding through Visual Feedback}
Upon trying system 1, participants highly valued the multimodal feedback provided which combines visual, audio, and textual elements to enhance clarity. 
A key advantage of the system is its ability to verify input through pictures, audio, and corrected spelling via speech-to-text. AP3 emphasized the importance of combining these methods to prevent errors, especially with homophones like \textit{“Sunday”} and \textit{“Sundae.”} When the system mistakenly generated images of a \textit{“Sundae”} for her speech input \textit{“Sunday,”} she noted an unexpected benefit: \textit{“Without this system and this mistake, I wouldn’t have realized such misunderstandings could happen. Even though it was wrong, it helped me clarify that I want ‘Sunday,’ not ‘sundae.’ Without it, the condition will be worse. That’s why we need it.”} While acknowledging that multiple verification steps can be tiring, she stressed their necessity: \textit{“Many steps is tiring, but necessary. Without it, will be worse.”}

Furthermore, visual aids could help reduce ambiguity and enhance mutual understanding. AP3 noted that images clarify meaning, especially when words have multiple interpretations. She explained, \textit{“If you say ‘apple,’ but you don’t have a photo to point to, people might think of an apple as a fruit, not in a pie. That’s why you need a photo to make sure everyone understands what you mean.”} Furthermore, AP7 described how her aphasia and dysarthria often lead to misunderstandings, particularly when she is nervous or tired. She shared, \textit{“Sometimes my speech gets in the way and I can’t speak. It happens at the most inconvenient times. I don’t know what’s going to come out of my mouth until I say it.”} Previously, she relied heavily on her husband to clarify her speech, which reduced her independence. She noted that the system could help her communicate on her own: \textit{“I think it’s going to be useful because you’re trying to track something, but you can’t say the word. We gotta speak for ourselves, you know. Because I say a lot to my husband. Can you tell him what I’m thinking? I know the story, but I can’t get it out of my mouth.”} (note: We  cut off empty words from this quote for clarity.)

The AI-generated images could potentially ensure all participants in a conversation stay aligned, even those unfamiliar with a PWA’s speech patterns. AP3 shared an example from a restaurant situation: \textit{“To make sure the team knows what’s going on. Like here—banana. If you’re thinking fruit or apple, people might not understand. So you need a photo of a banana, press the three banana options, and everybody will understand it’s banana.”} While long-term communication partners may understand PWAs more easily, strangers often struggle, so system 1 could be helpful then.

\subsubsection{Reduced Trust from Misalignment with AI}

A major challenge in building trust in AI lies in the misalignment between users’ intentions and the AI’s generated output. This misalignment stems from three key factors: AI instability, PWAs’ speech disorders, and speech recognition challenges.

The first factor, AI instability, occurs when the AI generates outputs that deviate from the user’s intended input, even if there is some surface-level association. For example, AP7 tried to verify the phrase \textit{“chocolate and a soup,”} but the system interpreted it as \textit{“chocolate in a soup,”} generating an incorrect image of a chocolate drink. She expressed frustration, exclaiming, \textit{“Oh, my God.”} In other cases, the AI failed to produce even loosely related results. When AP3 input \textit{“I want to go to my restaurant on Sunday for brunch,”} the AI generated an image of a picnic in a meadow. Similarly, AP8 input \textit{“soup,”} and the system produced an image of a book labeled \textit{“noun.”} Such errors undermine trust in the AI’s ability to accurately interpret user intentions.

The second factor involves the combined challenges of PWAs’ speech disorders—such as word retrieval difficulties, phonetic errors, and semantic confusion—and speech recognition limitations. AP1, struggling with word retrieval, spent five minutes attempting to verify a word but was unable to recall it. He pronounced \textit{“sheet”} but realized it was not the intended word, even after receiving related suggestions from system 1 like \textit{“seat, blanket, and pillow.”} None of these matched his intended meaning. In another instance, phonetic errors caused him to input \textit{“apple”} using his voice, which the system misinterpreted as \textit{“asshole,”} leading to frustration. Similarly, AP9 tried to verify \textit{“olives,”} but the system registered it as \textit{“always,”} resulting in an entirely different and incorrect output.

Despite these errors, users remained optimistic about the system’s potential for improvement. AP1, AP7 and AP8 all remarked: \textit{“It will be better with time.”}

\subsubsection{Reduced Grounding from Timing Issues}

A significant challenge highlighted by participants was the difficulty of using the system effectively in real-time group conversations. Although designed for such interactions, timing constraints often disrupted the flow of group discussions, making it impractical in such settings.

AP9 explained that why the system might not fit seamlessly into casual family conversations: \textit{“It’ll be hard to use when I’m with my family and talking with my family because I don’t… I don’t have a phone all the time to type something into or whatever. I mean, like, when we’re just talking about different things. Right? I’m not gonna go…look at my computer and check it, you know, we’re just having fun talking.”}

Similarly, AP3 noted that the system may be more suitable for one-on-one interactions than group settings:
\textit{“With a group, you have many people to understand, compared to just one. If I’m talking to one person, I can say, ‘I don’t understand,’ or ‘Can you help me?’ With one person, you can find a good word together. But in a group, like a family birthday, if I’m talking, who’s gonna stop and say, ‘This is good or bad’? They’ll just move on to the next person. In a group, there are too many people for one person. Nobody’s gonna stop and talk to you—they’ll just go on to the next thing.”}

\subsubsection{Suggestions to Enhance Accessibility}

Participants provided several suggestions to improve the system’s accessibility and usability, focusing on context-aware alternatives and visual transition experiences.

A key recommendation was to enhance context-aware alternatives by generating more related terms and pictures within the same category when the exact word cannot be recalled. Due to word retrieval difficulties, PWAs often rely on related words to convey their meaning. AP9 suggested that the system could offer multiple options within a category, such as various types of cakes: \textit{“The system can provide more cake options like cheesecake or strawberry cake. Because some aphasia cannot remember the exact one.”} AP7 shared a similar need, recounting how she input \textit{“sandwich”} and the system generated a picture of a generic sandwich. However, she selected \textit{“No”} because she was expecting more specific options, such as a veggie sandwich. Instead, the system generated unrelated images of \textit{“Sandwich, Wrap, and Burger.”} She commented: \textit{“Sometimes I wanna order a sandwich, and I don’t have the words here, right? Let me see what I can pull up. It’s… it’s from Panera. It’s a something, you know. It’s a veggie sandwich. But it’s a separate word, though.”} Despite these issues, AP7 appreciated the system’s ability to generate images of general categories, which helped her identify specific items like \textit{“banana”} from a general fruit picture.

Another suggestion was to improve the visual transition experience during image generation. Participants found the current spinning icon indicating processing time awkward. AP1 described it as \textit{“weird,”} saying: \textit{“The clock is going just a little around it around, and then it’s apple or pear. It’s just a little weird for me. Just the clock spinning, and then apple, or whatever it is, not an apple, or whatever it is. All of a sudden.”} He proposed making the image generation process smoother and more predictable to enhance the overall user experience.

\subsection{Results - System 2: Generate Sentences from keywords}

System 2 helps PWAs generate grammatically accurate and semantically complete sentences using keywords they can recall for three purposes: making requests, asking questions, or stating facts.

\subsubsection{Current Strategies for Constructing Full Sentences}
Forming full sentences is \textit{"very hard"} (AP8) for PWAs and remains a constant challenge \textit{"all the time"} (AP10), with recovery periods extending to\textit{ "six months"} for basic sentences (AP3). Our participants employ two primary strategies: therapist-taught techniques and external assistance. Therapeutic methods include tapping for word cueing (AP7) and structured prompt questions like "who, what, where, why" (AP3). For external support, PWAs rely on pre-writing techniques (AP1, AP9), often having family members review for clarity (AP7, AP9).

\subsubsection{Enhanced Trust in AI from High Autonomy}

Users expressed strong trust in System 2 and praised its effectiveness in helping with sentence construction. AP3 described it as the \textit{“best”} system, and also noted that mastering each generated sentence would take significant time: \textit{“We’ll look at that for maybe 6 months to make the sentence correct.”} AP8, who typically responds with single words or interjections like \textit{“oh”} and \textit{“yes,”} was excited when the system generated a full sentence from his keywords, saying: \textit{“Beautiful… Great sentences for words, man.”}

Users may have highly rated this system in terms of trust because of its ability to offer multiple sentence options, giving users control and flexibility. AP9 appreciated being able to adjust the suggestions: \textit{“I can change the words up too.”} AP10 valued having \textit{“sentences as examples”} that he could \textit{“see and change”} and the system’s multiple options reassured him: \textit{“No worries about autonomy.”} Similarly, other participants did not feel relying on AI would affect independence. Instead, our participants saw the system as a tool to improve social communication. AP3 explained: \textit{“I don’t think it will impair my independence because the computer is just helping to make sure the other person understands.”} AP8, who cannot independently form full sentences, was happy to use the system for social purposes: \textit{“It’s a social! I’m worried, but it’s to me… I’m happy.”}

Users also valued the system’s multiple sentence options as a backup when their efforts fell short. AP9 used the system to ask her sister to play board games. While she can usually form sentences independently, the system provided additional ideas. She explained: \textit{“I like to try things myself all the time. But I like this because it’s something else I can use in case it doesn’t come out right for me. All those questions are things that I would ask her, too. Sometimes you can’t come up with everything.”}

\subsubsection{Enhanced Grounding through ‘Three Intentions’ Feature}

Six participants highly valued the “three intentions” feature for its ability to better achieve context grounding with conversation partners (AP9) and \textit{"more efficiently"} identify targeted sentences (AP11), as it made it easier for others to understand their intentions. Among the three intentions, \textbf{“asking questions”} was considered the most useful by AP3, AP8, AP9, and AP7. Participants often use this intention for tasks like \textit{“asking about my appointments”} (AP3) or asking questions they hadn’t initially considered. As AP9 noted: \textit{“Sometimes you can’t come up with everything.”} AP7 further highlighted its value for \textit{“seeking external information and help.”}

The second most valued intention was \textbf{“stating facts,”} which reduces the burden of explaining oneself. AP7 shared an example of a voting discussion where this intention could have helped: \textit{“It would have been good for…about who I already voted for. Why did you vote for them? I think this is why I voted for her. But explaining myself is a burden, especially when they have a lot to say… Sometimes I don’t remember the whole story… I don’t know how to explain myself.”}

AP3 favored the \textbf{“request”} function, as she frequently needs to ask for accessibility accommodations in public spaces due to her reliance on a wheelchair. She explained how this feature could assist when visiting restaurants or hospitals: \textit{“I have a wheelchair, so I need wide access, no stairs. If I’m going to the ticket counter, I also need a seat. But asking with the computer could help, because people need to know. It helps explain disabilities clearly.”} (Paraphrased by cutting off empty words.)

\subsubsection{Reduced Trust from Misalignment and Aphasia Itself}

Misalignment between generated content and user intent also occurred in System 2. For instance, AP7 wanted to ask a colleague about purchasing a ticket and input the keywords “employee, date, theater.” However, the system provided irrelevant suggestions such as: “Is the employee scheduled to work at the theater today?” and “Do you need help finding the nearest movie theater location?” This left her frustrated. She noted that such mismatches could make her even more anxious, particularly when she is already nervous in public settings. However, AP9 acknowledged that mismatch issues might stem from challenges posed by aphasia itself. She described how her nervousness often led her to repeat words, which contributed to errors in speech recognition when using the system. She explained: \textit{“It’s probably me, too, again, is aphasia. Because I want to speak into it. It doesn’t get everything I say, because I also said, too. It takes in that, too. That’s my fault. I keep saying because maybe I’m nervous, maybe, or something I don’t know.”}

\subsubsection{Suggestions to Enhance Accessibility }

Additionally, AP3 suggested that for real-time conversations, the system should present an opening statement could help set the stage for effective interaction like: \textit{“Hi! I had a stroke. Can you be slower?”} She explained how this could prevent misunderstandings, especially with strangers who might otherwise misinterpret speech difficulties as signs of intoxication or erratic behavior: \textit{“For example, a waitress or a bus driver. They’ve never seen you before or know what you’re talking about. You say, ‘Hi, I had a stroke. Can you help me?’ Otherwise, the bus driver ’ll think you’re drunk, and it will just go away.”} 

Another common suggestion to improve accessibility was improving brevity in generated sentences. Although we prompted the system to generate simple sentences with a ten-word limit, participants often still felt it was too complex. AP3, AP9, and AP10 emphasized the importance of concise sentences, as long sentences overwhelm both speakers and listeners. AP3 explained: \textit{“Listeners will feel frustrated with long sentences.”}

To further enhance clarity in real-time conversation, AP1 suggested generating images related to sentences to aid communication, as he often relies on visual aids during conversations: \textit{“Well, the words and the pictures, or whatever they are, together, that’s fine.”}

Participants also suggested several potential contexts for using the system. AP1 and AP9 envisioned it being helpful for \textit{“public speaking conferences, church, and parties.”} AP8 and AP10 found it particularly suited for \textit{“email and messaging,”} as both rely heavily on written communication. AP10 also recommended integrating the system with \textit{“real-time messaging apps”} for everyday use. AP9 highlighted its value for \textit{“work-related tasks”} and as a tool for \textit{“students with aphasia.”} AP7 proposed using it during \textit{“grocery shopping”} to assist with communication, as her articulation disorder often makes her difficult to understand. However, she noted that regular use would depend on the \textit{“system’s reliability.”}

\subsection{Result - System 3: Check Whole Sentences}

System 3 helps PWAs for future conversations by identifying spelling, completeness, and contextual errors in text while providing multiple alternatives to enhance expression.

\subsubsection{Current Strategy to Ensure Writing Correctness}

PWAs experience writing challenges due to grammar (AP8, AP10), word retrieval (AP1), and phonetic disorders (AP9). Spontaneous writing is particularly \textit{"hard"}, requiring \textit{"a lot [of] time to prepare"} (AP7). For accuracy, our participants utilize multiple support systems: reference materials (AP3), expert guidance from therapists and doctors (AP3), family review (AP7, AP9), digital tools like Grammarly (AP8, AP10), and speech recognition with playback validation (AP1).

\subsubsection{Enhanced Grounding through Feedback Mechanisms}

System 3’s multimodal feedback—\textit{“reading, listening, and error correction”} (AP10)—helped participants improve sentence clarity and correctness. AP8 found it more specific to aphasia and useful than \textit{“Grammarly,”} even ranking it first in his notes, writing: \textit{“1. [anonymous name](researcher’s name), 2. Grammarly.”} He eagerly requested a link to start using the system daily.

The system’s multiple options also enriched expression diversity. AP1, who previously struggled to find varied adjectives, tested the system by inputting “happy for daughter’s wedding.” When the system suggested alternatives like “ceremony, marriage, celebration,” he selected “celebration” and remarked it was \textit{“really good and useful”} for enriching his expression.

The system’s high speech recognition accuracy brought unexpected benefits. Although designed for manual error correction, the AI model (OpenAI’s Whisper) automatically transcribed and corrected input errors, reducing users’ effort. AP9 was surprised at the system’s accuracy compared to other tools, stating: \textit{“I’m surprised because sometimes when I say something, it doesn’t come out right at all.”} She appreciated how it significantly minimized the need for manual corrections, easing the writing burden.

\subsubsection{Reduced Trust from AI’s Inability to Address Contextual Nuances}

A key challenge lies in the system’s inability to fully account for contextual nuances, which are particularly critical for PWAs. Due to aphasia, PWAs often recall the related words instead of the exact one they intend. For example, PWAs might want to say \textit{“doctor”} but can only recall \textit{“nurse.”} In such cases, the system should consider associations and broader contextual cues to provide meaningful suggestions. AP8 illustrated this issue when he wrote \textit{“I went to Balti”} (intending \textit{“Baltimore”}) but received unrelated suggestions like \textit{“Lahore, Gilgit, and Karachi.”} These alternatives were irrelevant, and he expressed a preference for contextually appropriate suggestions such as \textit{“Virginia”} or \textit{“New York.”}

Failing to consider the full context can still lead to unhelpful suggestions even when the system accounts for related words. AP7 wanted to thank her friend for treating her to dinner and input: \textit{“Michelle thnk you the diner last niht,”} expecting the corrected sentence to be \textit{“Michelle, thank you for treating me to dinner last night.”} Instead, the system offered suggestions related to dining locations like \textit{“restaurant, cafe, eatery,”} resulting in a confusing output: \textit{“Michelle thank you at the diner last night.”}

Another issue involves the vocabulary suggested by the AI, which, while technically correct, can lead to misinterpretation if the lexical nuances are not considered. AP3 shared an example where she input \textit{“I fel hapy for baseball,”} and the system suggested corrections for \textit{“fel”} as \textit{“feel”} or \textit{“fall.”} She criticized the output, saying: \textit{“Oh, no, it’s feel, not fall. If I’m falling, people will send an ambulance and go to the ER. I don’t want that. Even the word ‘feel’ is not good. People would understand ‘am’ better than feel because ‘feel’ makes people think of falling. That’s not a good word. A better word would be ‘I am.’”} (Paraphrased by cutting off empty words.)

These examples highlight the need for the system to better account for contextual nuances, associations, and the interpretive impact of its corrections on both PWAs and their communication partners.

\subsubsection{Suggestions to Enhance Accessibility}

Participants suggested ways to create a more accessible user experience, with four participants (AP3, AP1, AP8, AP4) emphasizing the need for \textit{“shorter, clearer”} sentences. The system initially generated long paragraphs because it processed speech into multiple lines without reorganizing for clarity, which many participants found frustrating. AP1 specifically disliked multi-line outputs, stating: \textit{“One line is awesome. Two is… a little weird because of one, and then two at the same time. Five or six lines… then it’s really not good. It’s absolutely too much.”}

AP3 suggested the system should remove \textit{“too long or unnecessary content”} and refine speech input into \textit{“clear and logical”} sentences. She stressed that concise content is essential for ensuring the listener remains patient and understands the intended message. She explained:

\textit{“Because the first time you talk, you do too much. Then you look at the writing and take things out, so for an email or letter, it’s now just short. People can look at a sentence and understand better. If I talk too much, like right now—talking about baseball tonight, the Dodgers, Disneyland, and pizza—you’re gonna get out of here. You don’t understand, and you won’t respond because I was talking too much and too fast. Then how can you come back to me? You can’t go back to how you were before.”} (Paraphrased by cutting off empty words)

These suggestions highlight the potential for the system to go further by reorganizing text into clearer, more concise, and well-structured outputs. This improvement could enhance accessibility, making it easier for both PWAs and their conversation partners to read and comprehend the content effectively.

\subsection{Result - System 4: Share Meaningful Experience}

System 4 helps PWAs prepare future conversations for experience sharing by generating highlights from user-uploaded photos and guided conversations.

\subsubsection{Current Strategy for Experience Sharing}

PWAs face significant cognitive challenges when sharing past experiences, particularly in organizing detailed narratives. Like AP7 explained: "\textit{can’t get in all details that I want to}". Our participants employ several coping strategies: using structured introductions and simple sentences (AP3), preparing with written tools (AP9), and relying on brief statements with family support for elaboration (AP7). While participants desire technology to enhance their narrative confidence (AP7), current tools present challenges, such as difficulty transferring content between platforms (AP1).

\subsubsection{Enhanced Trust through Personalized Summaries and Multimodal Support }

Participants identified several factors that build trust in AI, including personalization, accurate reproduction of details, and refining fragmented speech.

Summaries that reflected participants’ unique experiences were particularly valued over generic descriptions. For instance, AP3 highlighted how vivid and personalized details enhanced her engagement, citing an example of a beach trip summary: \textit{“I loved the white sandy beaches in Cuba. They were so beautiful and perfect for a resort vacation. The sand was as white as sugar, creating a picturesque setting for relaxation.”} She emphasized that details like \textit{“sand sugar”} made the experience feel personal, adding, \textit{“I prefer something personal, not just a normal description.”} Similarly, AP9 praised the system for capturing specific details, noting, \textit{“The detail in the summary was great because it used the proper words. It was almost exactly what I want to write about in my book.”}


\begin{figure}[h]
  \centering
  \includegraphics[width=0.9\linewidth]{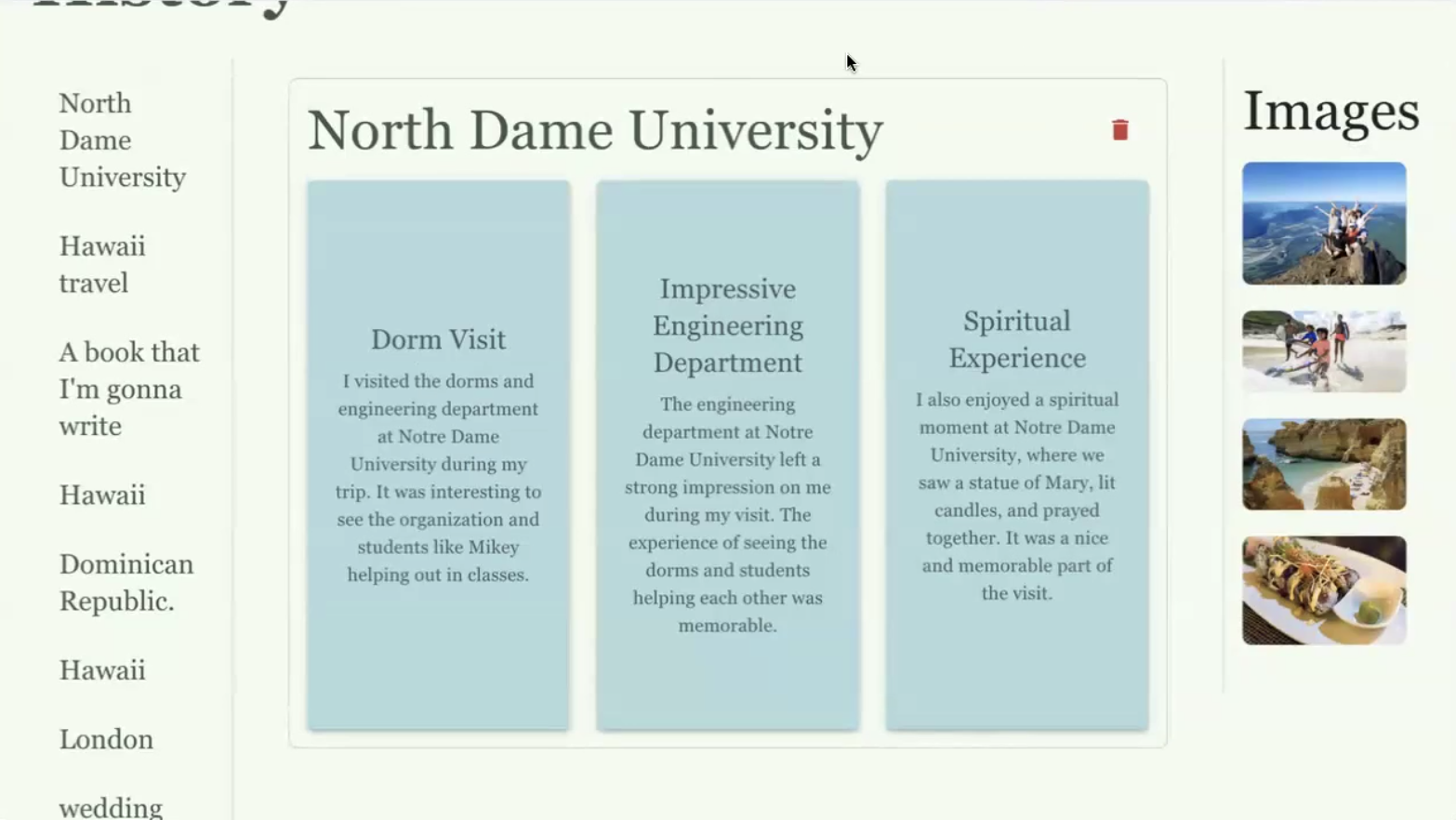}
\caption{An example of a generated diary summary from AP10, a participant with severe aphasia.}
  \label{fig:summary}
  \Description{A screenshot of System 4, showing three summaries from participants' inputs. The screenshot also displays images as visual aids.}
\end{figure}

The system’s ability to refine fragmented speech into clear and structured summaries was another key advantage. AP10 shared an example of his initial, disjointed input: \textit{“Mary! There’s a statue, I mean it’s…it’s nice! And we were pictures. Oh, and we had to…candles…you know…candles, and we’re…not a fire, but candles…and praying.”} The system transformed this into a coherent summary: \textit{“Spiritual Experience: I enjoyed a moment at Notre Dame University, where we saw a statue of Mary, lit candles, and prayed together. It was a memorable part of the visit.”} AP10 appreciated this, stating, \textit{“I was messing up, but they helped me organize it. Now it’s nice.”} AP9, who is writing a book about her daughter’s recovery, also found the system helpful, saying it encouraged her to speak more: \textit{“Usually I don’t speak much, but for this, I spoke a lot. Even though I spoke a lot, it made it a better sentence. The summary is really good.”}

Participants also valued System 4’s integration of speech, visual, and audio channels, which enhanced accessibility and clarity. AP3 explained how combining these sensory channels 
 improved communication: \textit{“You have three things…talking, visual, your photo, and audio, so people can hear what is going on. Your eye, mouth, and ear together…make sure you’re clear.”} She also emphasized the importance of visual aids: \textit{“You have to talk for sure…but having a photo helps more. They can understand better with the eyes.”}

\subsubsection{Reduced Trust Due to Irrelevance and Deviation from Original Intent}

Participants shared concerns about how well the AI understood their input. Key issues included mistakes in capturing what they meant, prompts that were irrelevant or repetitive, and summaries that didn’t match the right tone or context.

Participants felt dissatisfied when the system failed to capture their intended meaning accurately. For example, AP1’s input, \textit{“I go to London for to eat, and it is awesome,”} was misinterpreted as \textit{“I go to London for Eid, and it is awesome.”} This resulted in an irrelevant summary: \textit{“I go to London for Eid celebrations. And it’s a memorable experience celebrating in London.”} He described the output as \textit{“not good at all”} and \textit{“weird.”} However, a revised summary—\textit{“I go to London, and it was fantastic. I had a great time with one highlight being the amazing sushi”}—was far more satisfactory. He called it \textit{“a really good sentence… not just okay or bad, but really good.”}

The relevance of prompts was another critical issue. Irrelevant or repetitive prompts frustrated participants, reducing their trust in the system. AP1 labeled irrelevant prompts as \textit{“bad prompts,”} while AP9 highlighted the negative impact of repeated questions: \textit{“It just repeated something at the end. Repeat questions are not good.”}

Flexibility in language style also emerged as an important factor. AP3 shared an example involving her input about her son’s wedding: \textit{“Oh, the groom was laughing. He was in a beach, so he only had shorts. He didn’t have a tux.”} The system generated the summary: \textit{“The groom opted for a relaxed beach look with shorts and no tux, enjoying the fun atmosphere of the beach wedding.”} While she appreciated the summary’s accuracy, she noted that overly formal expressions could be unsuitable in informal contexts. She explained, \textit{“If you’re talking about no tux and shorts, people might not like it. This is only for personal use, not for a group or formal setting.”}

The last issue was the lack of clear guidelines on how many interaction rounds were needed to complete a prompt conversation. This ambiguity often led to frustration when the process felt unclear: \textit{“I didn’t know how many rounds to share, and by the end, I felt frustrated} (AP3).” AP3 recommended providing a clear timeframe, such as \textit{“five rounds of talk to record,”} to make the process more predictable.

\subsubsection{Usefulness Varies by Aphasia Severity and Goals}

System 4’s effectiveness depends largely on the user’s aphasia severity and individual goals.

Participants believed the system could help them express themselves more effectively, especially when interacting with strangers. AP7 shared, \textit{“It will help me express myself more because sometimes I don’t talk much when I’m outside. I think it would help.”} Similarly, AP9 appreciated its ability to encourage expression, highlighting its versatility: \textit{“Sometimes I want to talk but can’t find the words. If I can say what I want and someone summarizes it, that’d be amazing. I can use it all the time in every part of my life.”}

However, concerns were raised about the system’s usability for individuals with more severe aphasia. While some found it helpful for basic conversations, those with severe symptoms might face challenges and frustration. AP9 explained, \textit{“It depends on your aphasia. Before, it was hard for me to remember things or say even brief sentences. But now I’m better with aphasia, so it works for me. For people with severe aphasia, they might only say one word or not respond at all.”}

Additionally, some participants felt the system was less applicable for daily use, particularly for those focusing on reading and writing recovery. AP10 rated it lower for everyday utility, stating, \textit{“Because I had this stroke, I’m focused on reading and writing. That’s my priority. This system is nice for events like weddings or pictures, but it’s not my priority.”} (Paraphrased by cutting off empty words.)

\section{Discussion and Future Work}


Semantic and phonetic disorders present significant challenges for PWAs, yet technical solutions for validating information remain underexplored. AP2 and AP7 highlighted the emotional burden of misspeaking. To address this, System 1 uses AI-generated images and related words to help PWAs validate their expressions and find alternatives, assisting without fully replacing user input. This approach reduces dependence on automated phrase generation~\cite{shen2022kwickchat}.

For System 2, we introduced three intent buttons inspired by common conversational goals~\cite{valencia2024compa}. This feature, a favorite among five participants, helps PWAs find their intended expressions more efficiently and ensures AI-generated content aligns with user intent. However, AP10 suggested adding more customizable intent options to expand users’ expressive capabilities~\cite{valencia2024compa}.

Recent AAC research has also emphasized the importance of social factors, such as agency~\cite{ibrahim2018design, valencia2020conversational} and relationship maintenance~\cite{dai2022designing}. Our designs aim to support PWAs in overcoming not only linguistic challenges but also emotional barriers in social contexts. Participants stressed the importance of reaching a shared understanding with strangers and reducing stereotypes about aphasia. As AP2 noted, greater public awareness and acceptance can foster interdependence, breaking down societal barriers and encouraging PWAs to engage socially. 

In this work, we also examine two important phenomena, trust in AI-enhanced systems and grounding (mutual understanding) between the system and the user. We learned that for PWA, trust is built when systems accurately capture user intent, provide flexible options for autonomy and expression, and offer contextually appropriate suggestions while refining fragmented speech into clear outputs. Grounding is supported by visual aids and multimodal feedback, which improve clarity, reduce ambiguity, and promote independent communication. In contrast, grounding is undermined by misalignment between intent and AI outputs, irrelevant suggestions, and timing issues in real-time conversations. These takeaways can contribute to the design of future AAC systems.

\subsection{Reflecting on Existing Work}

Previous studies have focused on helping PWAs retrieve vocabulary ~\cite{kim2009context, nikolova2010click, nikolovaBetterVocabulariesAssistive2009}, often using algorithms to predict or generate words based on context~\cite{kane2012we,kane2017let}. In contrast, our designs prioritize sentence-based interactions to enable more effective conversations. Research in AAC~\cite{todman2000rate, wisenburn2009participant} shows that sentence-based systems allow faster engagement in conversations. Reflecting on this, three of our four systems emphasized sentence construction over word retrieval.

While recent advancements in AI-enhanced AAC suggest context-aware phrases can improve communication 
~\cite{valencia2023less}, they may not fully meet user needs. PWAs value personal expression and unique communication styles, as seen in System 4, where participants preferred personalized over generic text generation. Future systems could leverage large language models (LLMs)~\cite{cai2022context,cai2023speakfaster,shen2022kwickchat,valencia2023less} to better adapt to users’ contexts, communication styles, and preferences, addressing their specific needs more effectively.

\subsection{Reflections on Design Probe Study Using Real Data and Models}

The heterogeneity of aphasia makes it essential to avoid assuming AAC user preferences. To minimize biases from designers’ “ableism,” which can lead to neglecting disabled individuals’ perspectives~\cite{hofmann2020living}, our study used live prototypes and a participatory exploration approach~\cite{zimmermanResearchDesignMethod2007}. This allowed us to identify shortcomings in existing features and interaction mechanisms and collaborate with participants to generate ideas for future AI-driven AAC systems.

For example, System 1 was initially designed to use AI-generated images to verify expressions and address semantic errors, such as confusing “nurse” with “doctor.” However, testing revealed that participants preferred using images as reminders for word retrieval or as scaffolds to build shared understanding in conversations rather than just for verification. Image validation was most helpful in two scenarios: for users with semantic disorders to confirm their intended meanings and for users with speech disorders to clarify their spoken messages and reduce ambiguity. These findings suggest that future designs should explore image generation as a real-time tool for enhancing shared understanding.

We also assumed PWAs would favor speech input due to motor disabilities, but this proved challenging for participants with articulation disorders. Others, experiencing post-aphasia anxiety, preferred typing over speech. This highlights the need for context-aware input options tailored to specific subgroups or scenarios. For example, many participants expressed interest in using the system while driving, emphasizing the importance of situational adaptability.

These findings underscore the need to critically reflect on design assumptions and create tools that accommodate diverse user needs without introducing new challenges or risks.

\subsection{Heterogeneity of Aphasia as a Key Consideration in AAC Design}

The wide range of aphasic symptoms makes it difficult to design AAC systems that accommodate all users without excluding some. No single system can provide broad inclusivity, as accessibility needs vary greatly among PWAs. System evaluations highlighted this variability, showing polarized feedback. For example, some participants found System 4 ideal, particularly those with good conversational skills but significant cognitive load when organizing complex language. Others, prioritizing basic literacy skills, found it less useful due to their lower demand for event recording and sharing.

As Hofmann et al. noted, ~\cite{hofmann2020living} current technologies struggle to address these conflicting needs or balance accessibility for diverse users. Our research found that PWAs are aware of this challenge and often consider the needs of others with different symptoms when evaluating systems. This underscores a major challenge in our AAC designs—they fail to support a collaborative, inclusive approach. Addressing this requires making user diversity a core principle, which, though challenging, is essential for creating impactful, inclusive designs.

\subsection{Impact of AI on PWAs}

Using AI systems as probes revealed that AI could have an adverse impact on PWAs, influencing their expression style, communication speed, tolerance for uncertainty, and ability to maintain intent, a concern mentioned in other work~\cite{valencia2023less}.

First, AI tools, as proposed in our designs, may change how PWAs express themselves. Unlike traditional AAC devices that rely on fixed words and images, AI generates flexible outputs, giving users greater control over their communication but also priming new ideas. Contrary to concerns that AI might reduce autonomy, our participants found it could  help them express their needs more freely. However, the unpredictability of AI-generated content introduces risks, such as increased system errors and hidden bias~\cite{mack2024they, glazko2024identifying}, that may also create more user frustration and confusion. 

Second, AI impacts communication speed. Its variability allows users to regenerate outputs multiple times, providing flexibility but also adding cognitive burden. PWAs must interpret changing outputs and make decisions quickly, which can be challenging for those with language difficulties and cognitive fatigue. If AI outputs do not align with user intent, re-entering inputs further slows communication.

Third, AI affects users’ tolerance for mismatches. While flexibility enables the exploration of different expressions, users may become less tolerant of AI-generated content that deviates from their intent. For instance, during System 4 evaluations, users expressed greater dissatisfaction with AI errors than with their own communication mistakes. When asked if AI understood their intent or if they trusted it, responses varied significantly due to personal preferences and experiences.

Finally, AI might influence users’ ability to maintain their original intent. As AI improves and its errors become subtler~\cite{google_pair_2021}, users may trust its outputs even when they are not entirely accurate. The natural language or visual style of AI-generated content can lead users to unknowingly rely on it, potentially undermining their autonomy over time.

These findings reveal that while AI enhances expression for PWAs, it also introduces challenges in autonomy, efficiency, and bias. Balancing these risks and benefits is crucial for designing reliable AAC to support PWAs' needs.

\subsection{Managing Time Commitment and Fatigue in Remote User Testing}

Managing cognitive load is essential for inclusive design during system testing. Despite efforts informed by aphasia-related research, speech therapy practices, and methods from online aphasia support groups, challenges related to time, effort, and fatigue persisted. Research objectives required careful consideration of how to allocate PWAs’ limited energy between user testing and interactive evaluation.

Remote testing via Zoom introduced significant accessibility barriers for PWAs. Many participants struggled with language comprehension, finding it difficult to understand interface terminology~\cite{chen2023exploring} and remember tutorial steps. Technical operation of personal computers posed further challenges, as most participants, being older adults, were unfamiliar with technology. Physical limitations, as mentioned in previous work~\cite{williamsDesigningConversationCues2015}, compounded these challenges by impeding touchscreen interactions.

To address these challenges, we shifted from traditional user testing to using systems as “probes” to explore PWA needs. This approach reduced cognitive load, enabling participants to focus their feedback on the systems. However, since participants did not directly operate the systems, some feedback may have been overly optimistic, highlighting a trade-off between reducing cognitive strain and ensuring authentic evaluations. Future designs must balance operational accessibility with genuine user engagement to enhance both testing reliability and participant experience.

\section{Conclusion}

In this work, we sought to understand how to leverage AI's specific qualities to support PWAs' real-time communication and preparation for future conversations. Therefore, we developed four AAC prototypes integrating various AI technologies as a design probe. We used these prototypes as design probes with 11 participants with aphasia to elicit conversations and reactions about the use of AI for communication support. We identified existing challenges in communication, impressions of potential design concepts, and desires for an ideal future of AI-enhanced AAC systems specific for users with aphasia.






\begin{acks}
To Robert, for the bagels and explaining CMYK and color spaces.
\end{acks}

\bibliographystyle{ACM-Reference-Format}
\bibliography{sample-base}


\end{document}